\documentclass{article}[11pt]
\usepackage{amssymb,latexsym,amsmath}
\usepackage[dvips]{graphicx}
\def\beq{\begin{equation}}
\def\eeq{\end{equation}}
\def\bea{\begin{eqnarray}}
\def\eea{\end{eqnarray}}
\def\nn{\nonumber}

\def\diag{\mathop{\rm diag}\nolimits}
\def\qdots{\mathinner{\mkern1mu\raise1pt\vbox{\kern7pt\hbox{.}}\mkern2mu
 \raise4pt\hbox{.}\mkern2mu\raise7pt\hbox{.}\mkern1mu}}

\def\sq#1{\big\lgroup#1\big\rgroup}
\def\s#1{\bigg\lgroup#1\bigg\rgroup}
\def\sl#1{\left\lgroup#1\right\rgroup}
\def\sB#1{\Bigg\lgroup#1\Bigg\rgroup}
\renewcommand{\atop}[2]{\genfrac{}{}{0pt}{}{#1}{#2}}

\begin{document}
\begin{center}
{\Large \bf
Quantum communication through Jaynes-Cummings-Hubbard arrays
}\\[5mm]
{\bf R.~Chakrabarti$^{1}$\footnote{E-mail: ranabir@imsc.res.in;
} and G.~Sreekumari$^{1, 2}$}\footnote{E-mail:
gsreekumari@imsc.res.in.}\\[1mm]
${}^{1}$Department of Theoretical Physics, University of Madras, Guindy Campus, Chennai 600 025,
India\\
${}^{2}$Department of Physics, Loyola College, Chennai 600 034, India.
\end{center}

\vskip 10mm



\begin{abstract}

We study the dynamics of an one dimensional array of Jaynes-Cummings-Hubbard system of arbitrary number of coupled cavities,
each containing a two level atom that interacts with a field mode. In particular, we consider propagation of a single excitation
quantum state for two different couplings of the photonic modes of the adjacent cavities, namely, a translation invariant closed
chain of uniformly coupled cavities, and also a linear chain with nonuniform parabolically varying intercavity coupling where the interaction
Hamiltonian is associated with the Jacobi matrix of the Krawtchouk polynomials. Using a description via the delocalized atomic and
field modes we observe that for a large detuning of
these two degrees of freedom atomic excitations propagate without populating the field modes, and {\it vice versa}. For the near-resonance scenario
between these modes the  atomic excitations, say, while propagating mix with the photonic states. In the context of the parabolic
coupling between photons of adjacent cavities an arbitrary element of the time-dependent correlation function between two arbitrary cavities may be
expressed in closed form for dominant values of the detuning parameter, when an exact transmission of the quantum state at pre-specified times is realized.

\end{abstract}

\section{Introduction}
\label{intro}
Recently considerable theoretical and experimental attention has been devoted to a class of models involving
coupled optical cavities where photons interact with embedded two level atoms. On the experimental side
these models have been facilitated  by experimental advance in photonic crystals \cite{AKSV2003}, optical
microcavities \cite{BTO2000}, and superconducting devices \cite{WSBFHMKGS2004}. These coupled cavity
structures have potential applications as quantum optical simulators \cite{ASB2007, IOK2008} of
condensed-matter phenomena making it feasible to study many-body effects such as phase transitions where the
particles of interest are photons. The coupling between cavities also provides a setting for the system being
considered as a wave guide in the context of distributed quantum information processing \cite{CEHM1999}. Allowing
for control and measurement at individual lattice sites proposals have been put forward for generating entangled
photonic states \cite{AB2007}, creation of cluster states \cite{AK2008, HBP2007}, and transfer of a quantum
state along an array of polaritonic qubits \cite{BAB2007}. Towards investigating the transmission of quantum states
in an one-dimensional coupled array of cavities much study has been done in the context of  Jaynes-Cummings-Hubbard
model \cite{GTCH2006}. The atom-cavity photon interaction is described by the well-understood  Jaynes-Cummings model \cite{JC1963}
that relies on rotating wave approximation. Tunneling of photons between
adjacent cavities is introduced via a hopping parameter. As the number of excitations in the system remain conserved
it is, from the point of view of propagation of quantum states, of importance to develop a detailed understanding
of the time evolution of single excitation states. Studying a system of two coupled cavities the authors of
Ref. \cite{OIK2008} introduced delocalized modes to reveal that the detuning parameters between the atoms and these modes
govern the dynamics of the propagation.  The possibility of control of the individual coupling constants between the
adjacent cavities has been investigated in \cite{MCHGH2009}. For the choice of a coupling that varies parabolically on the
cavity sites a dispersion-free propagation of the single excitation states has been observed \cite{MCHGH2009}.

\par

Following the procedure of \cite{OIK2008} here we study the propagation of atomic and photonic states between two
sites of an array of cavities with two different choices of coupling coefficients linking photons of adjacent cavities.
Generalizing the results of \cite{OIK2008} we first consider  a translation invariant closed chain of $N$ identical
cavities with periodic boundary condition where the coupling coefficient between the photons of nearest-neighbor
cavities is held uniform. Introducing the delocalized atomic and photonic
modes we follow a time-averaging procedure \cite{JJ2007, OIK2008} to perturbatively evaluate the effective Hamiltonian
for the three limiting cases:  $(\mathsf{i})$ dominant hopping parameter, $(\mathsf{ii})$ large detuning between the atomic and
photonic frequencies, and  $(\sf{iii})$ the case when the resonance between the atomic frequency and an eigenfrequency of a
delocalized photonic mode is realized. It is interesting to note that in terms of delocalized atomic coordinates a {\it diagonal}
spin-spin effective interaction that directly transfers energy without the intermediacy of a photonic process is introduced.

\par

Allowing for the precise control of the coupling constant between the adjacent sites the authors of Ref. \cite{MCHGH2009} introduced
the parabolic coupling scenario. In the context of one dimensional spin chains  \cite{B2003} it has been observed \cite{ACDE2004} that
the parabolic coupling between nearest-neighbor spins leads to exact transmission of a single excitation quantum state. In the limit of
a dominant hopping parameter between the photons of the adjacent sites perfect transfer of a quantum state at pre-specified times is
also realized in the present case. In the model studied here the atom-photon coupling at individual cavities introduces a
dispersive effect that prohibits perfect transfer of  a quantum state for large values of the said coupling. However, in
another limiting case when the detuning parameter is large compared to both the hopping parameter and the atom-cavity photon coupling
constant, exact propagation of the quantum state is also
obtained with an increment in the required time. For these cases the exact time-dependent correlation function of the quantum state
between two arbitrary cavities is obtained in a closed form. The plan of the paper is as follows. In Sec. \ref{cyclic} we discuss the case
of uniform coupling between the photons of adjacent cavities arranged periodically on a closed loop. Our discussion of the
case of parabolic coupling between the photons of adjacent cavities is contained in Sec. \ref{parabolic}.  Following this we conclude.

\section{Cyclic cavities with uniform coupling}
\label{cyclic}
\setcounter{equation}{0}
Here we consider a cyclic chain of $N$ identical cavities obeying periodic boundary condition, and each containing a two-level atom
that is coupled to a  localized photonic  mode modeled as a harmonic oscillator. The transition frequency $\varepsilon$
of the atoms and  the frequency $\Omega$ of the oscillators are held uniform over the chain.
The atom-cavity photon interaction is described by Jaynes-Cummings model with the adoption of rotating wave approximation.
The adjacent cavities are interlinked via photon hopping with uniform couplings between neighboring
cavities. The Hamiltonian may be expressed as a sum of the Jaynes-Cummings
Hamiltonians for identical cavities and a nearest-neighbor photon hopping term:
\beq
H = \sum_{j = 0}^{N-1} H_{j}^{\hbox{JC}} + H^{\hbox{hop}},
\label{H_tot}
\eeq
where the Jaynes-Cummings Hamiltonian for the $j$-th cavity  and the hopping term between
adjacent cavities, respectively, read:
\bea
H_{j}^{\hbox{JC}} &=& \varepsilon\, \sigma_{j}^{+} \sigma_{j}^{-} + \Omega \,a_{j}^{\dagger} a_{j}
+ \mathsf{g} \left(\sigma_{j}^{+} a_{j} + \sigma_{j}^{-} a_{j}^{\dagger}\right),\nn\\
H^{\hbox{hop}} &=& \kappa \sum_{j, k = 0}^{N-1} a_{j}^{\dagger}\, {\cal C}_{j, k}\, a_{k}, \qquad
{\cal C}_{j, k} = \delta_{j\,k+1} + \delta_{j+1\,k}.
\label{H_comp}
\eea
In the above expression the $N$-th and the $0$-th degrees of freedom are identified.
The standard commutation relations for the photonic and the atomic modes pertaining to
individual cavities are given below:
\beq
[a_{j}, a_{k}^{\dagger}] = \delta_{j k},\quad
[\sigma_{j}^{z}, \sigma_{k}^{\pm}] = \pm 2\, \delta_{j k}\, \sigma_{k}^{\pm},\quad
[\sigma_{j}^{+}, \sigma_{k}^{-}] = \delta_{j k}\, \sigma_{k}^{z}.
\label{loc_com}
\eeq
The operator ${\cal N}$ that represents the total number of atomic and photonic excitations
of the combined system commutes with the Hamiltonian (\ref{H_tot}):
\beq
{\cal N} = \sum_{j = 0}^{N-1} \left(\sigma_{j}^{+} \sigma_{j}^{-} + a_{j}^{\dagger} a_{j}\right),\qquad [H, {\cal N}] = 0.
\label{N_loc}
\eeq
In this work we focus on the time-evolution of single excitation states that may be arbitrary superpositions
of one excitation states of atoms and photons. The ground states of the photonic and atomic systems
may be listed as $|\mathbf{0}\rangle \equiv  |\{0_{j}\}\rangle,\;|\mathbf{G}\rangle \equiv  |\{g_{j}\}\rangle$
for $j = 0, 1,\ldots, N-1$. The photonic and atomic one excitation states localized in $j$-th cavity are created as follows:
\beq
a_{j}^{\dagger} |\mathbf{0}\rangle = |1_{j}\rangle, \qquad a_{j} |1_{k}\rangle = \delta_{j\, k}\;|\mathbf{0}\rangle,\qquad
\sigma_{j}^{+} |\mathbf{G}\rangle  = |e_{j}\rangle, \qquad \sigma_{j}^{-}|e_{k}\rangle = \delta_{j\, k}\;|\mathbf{G}\rangle.
\label{loc_st}
\eeq

\par

To diagonalize the photonic part of the Hamiltonian we need to introduce delocalized coordinates unitarily related to their
local analogs pertaining to a cavity. For $\omega$ being a root of unity the unitary transformation reads
\beq
U_{j k} = \frac{1}{\sqrt{N}} \omega^{j k}\;\;\forall j, k = 0, 1, \ldots, N - 1,
 \qquad \omega^{N} = 1 \;\;\;\Rightarrow \;\;\; \omega = \exp\left(2 \pi i/N\right),
\label{trns}
\eeq
and the unitarity constraint readily follows:
\beq
\frac{1}{N} \sum_{j = 0}^{N - 1} \omega^{j(k - \ell)} = \delta_{k \ell}\;\;\; \Rightarrow \;\;\; U U^{\dagger} = U^{\dagger}U = \mathbb{I}.
\label{unitary}
\eeq
We note that here and elsewhere in  this section we follow the $\hbox{mod}\, N$ arithmetic.  Now the delocalized photonic coordinates
obeying canonical commutation relations are given by
\beq
A_{j} = \frac{1}{\sqrt{N}} \sum_{k = 0}^{N - 1} \omega^{j k}\; a_{k}, \qquad
A_{j}^{\dagger} = \frac{1}{\sqrt{N}} \sum_{k = 0}^{N - 1} \omega^{- j k}\; a_{k}^{\dagger}\quad \Rightarrow \quad
[A_{j}, A_{k}^{\dagger}] = \delta_{j k}.
\label{A_deloc}
\eeq
It turns out that the atomic coordinates that couple with the delocalized photonic coordinates
in the reconstructed interaction Hamiltonian are also delocalized in nature. These collective spin variables are
introduced as discrete Fourier transforms
\beq
S_{j}^{\pm} = \frac{1}{\sqrt{N}} \sum_{k = 0}^{N - 1} \omega^{\mp j k}\; \sigma_{k}^{\pm}, \qquad
S_{j}^{z} = \frac{1}{N} \sum_{k = 0}^{N - 1} \omega^{- j k}\; \sigma_{k}^{z},\qquad
\left(S_{j}^{\pm}\right)^{\dagger} = S_{j}^{\mp},\qquad \left(S_{j}^{z}\right)^{\dagger} = S_{- j}^{z}
\label{S_deloc}
\eeq
obeying a closed algebra:
\beq
[S_{j}^{z}, S_{k}^{\pm}] = \pm \,\frac{2}{N}\; S_{k \pm j}^{\pm}, \qquad
[S_{j}^{+}, S_{k}^{-}] = S_{j - k}^{z}.
\label{S_com}
\eeq
The unitary operator (\ref{trns}) diagonalizes the coupling matrix between adjacent cavities given in (\ref{H_comp}):
\beq
\left(U\,{\cal C}\,U^{\dagger}\right)_{j k} = \left(\omega^{j} + \omega^{- j}\right)\,
\delta_{j k} = 2\, \cos\left(j\,\frac{2 \pi}{N}\right)\, \delta_{j k}.
\label{A-diag}
\eeq
Recasting the Hamiltonian (\ref{H_tot}) via the delocalized coordinates we obtain
\beq
H = H_{0} + H^{\hbox{int}},\qquad H_{0} = H_{0}^{\hbox{cavity}} + H_{0}^{\hbox{photon}},
\label{H0_decom}
\eeq
where
\beq
H_{0}^{\hbox{cavity}} = \varepsilon \;\sum_{j = 0}^{N - 1} S_{j}^{+} S_{j}^{-}, \qquad
H_{0}^{\hbox{photon}} = \sum_{j = 0}^{N - 1} \Omega_{j} A_{j}^{\dagger} A_{j}, \qquad
\Omega_{j} = \Omega + 2 \kappa \,\cos \left(j\,\frac{2 \pi}{N}\right).
\label{H0}
\eeq
For $N > 2$ the eigenfrequencies $\Omega_{j}$ of the delocalized modes  $A_{j}$ are degenerate:
$\Omega_{j} = \Omega_{N-j} \; \hbox{for}\; j \neq (N - j)\; \hbox{mod}\; N$. The non-degenerate
eigenfrequencies correspond to the `center-of-mass' mode $A_{0}$ and `alternating' mode
$A_{N/2}$ for even $N$. The delocalized coordinates introduced above maintains the
structure that reflects the rotating wave approximation for the atom-cavity photon interaction
term $H^{\hbox{int}}$ in (\ref{H0_decom}):
\beq
H^{\hbox{int}} = \mathsf{g}\;\sum_{j = 0}^{N-1} \left(S_{j}^{+} A_{j} + S_{j}^{-} A_{j}^{\dagger}\right).
\label{H_SA}
\eeq
The total number of excitation operator (\ref{N_loc}) now assumes the form
\beq
{\cal N} = \sum_{j = 0}^{N-1} \left(S_{j}^{+} S_{j}^{-} + A_{j}^{\dagger} A_{j}\right).
\label{N_deloc}
\eeq

\par

The delocalized photonic and atomic single-excitation states may be introduced via invertible
unitary transformation as follows:
\bea
|\tilde{1}_{j}\rangle &\equiv& A_{j}^{\dagger} |\mathbf{0}\rangle =
\frac{1}{\sqrt{N}} \sum_{k = 0}^{N - 1} \omega^{- j k} |1_{k}\rangle, \quad
A_{j} |\tilde{1}_{k}\rangle = \delta_{j k} |\mathbf{0}\rangle, \quad
\langle \tilde{1}_{j}|\tilde{1}_{k}\rangle = \delta_{j k},\nn\\
|\vartheta_{j}\rangle &\equiv& S_{j}^{+} |\mathbf{G}\rangle =
\frac{1}{\sqrt{N}} \sum_{k = 0}^{N - 1} \omega^{- j k} |e_{k}\rangle, \quad
S_{j}^{-} |\vartheta_{k}\rangle = \delta_{j k} |\mathbf{G}\rangle, \quad
\langle \vartheta_{j}|\vartheta_{k}\rangle = \delta_{j k}.
\label{deloc_st}
\eea
The localized and the delocalized one excitation states introduced in (\ref{loc_st}) and
(\ref{deloc_st}) are two sets of mutually unbiased bases:
\beq
\langle 1_{j}| \tilde{1}_{k}\rangle = \frac{1}{\sqrt{N}}\,\omega^{- j k},\qquad
\langle e_{j}| \vartheta_{k}\rangle = \frac{1}{\sqrt{N}}\,\omega^{- j k}\quad \Rightarrow \quad
\left|\langle 1_{j}| \tilde{1}_{k}\rangle\right| = \left|\langle e_{j}| \vartheta_{k}\rangle\right|
= \frac{1}{\sqrt{N}}.
\label{mub}
\eeq
The distinction between the localized and delocalized states may be summarized by introducing a unitary shift
operator that translates localized  photonic and atomic states through a single cavity. Specifically, it acts
on the single excitation states as follows:
\beq
\tau |1_{j}\rangle = |1_{j + 1}\rangle, \qquad \tau^{-1} |1_{j}\rangle = |1_{j - 1}\rangle, \qquad
\tau |e_{j}\rangle = |e_{j + 1}\rangle, \qquad \tau^{-1} |e_{j}\rangle = |e_{j - 1}\rangle\qquad
\tau^{\dagger} = \tau^{-1}.
\label{shift_loc}
\eeq
The construction (\ref{deloc_st}) makes it apparent that the delocalized states are invariant under
the action of the shift operator $\tau$ where the eigenvalues correspond to the root of unity phase angles:
\beq
\tau |\tilde{1}_{j}\rangle = \omega^{j} |\tilde{1}_{j}\rangle, \qquad
\tau |\vartheta_{j}\rangle = \omega^{j}|\vartheta_{j}\rangle.
\label{shift_deloc}
\eeq
The delocalized photonic and atomic operators introduced in (\ref{A_deloc}) and (\ref{S_deloc})
transform as follows:
\beq
\tau \,A_{j}\, \tau^{-1} = \omega^{-j}\,A_{j},\qquad
\tau \,A_{j}^{\dagger}\, \tau^{-1} = \omega^{j}\,A_{j}^{\dagger},\qquad
\tau \,S_{j}^{\pm}\, \tau^{-1} = \omega^{\pm j}\,S_{j}^{\pm}.
\label{shift_op}
\eeq
It is observed that the interaction Hamiltonian (\ref{H_SA}) is invariant under the shift transformation:
\beq
\tau \,H^{\hbox{int}} \,\tau^{-1} = H^{\hbox{int}}.
\label{tau_inv}
\eeq

\par

The most general one excitation state may be equivalently expressed as linear compositions of either the
localized or the delocalized basis states. These alternate expansions are listed, respectively, below:
\beq
|\Psi (t)\rangle = \sum_{j = 0}^{N-1}\;
\left(\mathsf{a}_{j}(t)\;|\mathbf{G}\rangle \otimes |1_{j}\rangle \;
+ \;\mathsf{b}_{j}(t)\;|e_{j}\rangle \otimes |\mathbf{0}\rangle\right),
\label{state_loc}
\eeq
\beq
|\Psi (t)\rangle = \sum_{j = 0}^{N-1}\;
\left(\alpha_{j}(t)\;|\mathbf{G}\rangle \otimes |\tilde{1}_{j}\rangle \;
+ \;\beta_{j}(t)\;|\vartheta_{j}\rangle \otimes |\mathbf{0}\rangle\right).
\label{state_deloc}
\eeq
Invertible unitary transformations interrelate the coefficients of the above
expansions (\ref{state_loc}) and (\ref{state_deloc}):
\beq
\alpha_{j}(t) = \frac{1}{\sqrt{N}}\,\sum_{k = 0}^{N-1}\; \omega^{j k}\,
\mathsf{a}_{k}(t),\qquad
\beta_{j}(t) = \frac{1}{\sqrt{N}}\,\sum_{k = 0}^{N-1}\; \omega^{j k}\,
\mathsf{b}_{k}(t),
\label{albe_exp}
\eeq
\beq
\mathsf{a}_{j}(t) = \frac{1}{\sqrt{N}}\,\sum_{k = 0}^{N-1}\; \omega^{- j k}\,
\alpha_{k}(t),\qquad
\mathsf{b}_{j}(t) = \frac{1}{\sqrt{N}}\,\sum_{k = 0}^{N-1}\; \omega^{- j k}\,
\beta_{k}(t).
\label{ab_exp}
\eeq
The action of the Hamiltonian (\ref{H0_decom}) on the two dimensional subspace
$\{|\mathbf{G}\rangle \otimes |\tilde{1}_{j}\rangle, \,
|\vartheta_{j}\rangle \otimes |\mathbf{0}\rangle\}$ reads
\bea
H\, |\mathbf{G}\rangle \otimes |\tilde{1}_{j}\rangle &=&
\Omega_{j} |\mathbf{G}\rangle \otimes |\tilde{1}_{j}\rangle
+ \mathbf{g} \,|\vartheta_{j}\rangle \otimes |\mathbf{0}\rangle, \nn\\
H\, |\vartheta_{j}\rangle \otimes |\mathbf{0}\rangle &=&
\varepsilon\, |\vartheta_{j}\rangle \otimes |\mathbf{0}\rangle +
\mathbf{g}\, |\mathbf{G}\rangle \otimes |\tilde{1}_{j}\rangle.
\label{H_act}
\eea
The Schr\"{o}dinger equation for the state $|\Psi (t)\rangle$
now block diagonalizes the one excitation subspace of the Hamiltonian (\ref{H0_decom}) in
$2 \times 2$ blocks with the $j$-th block given by
\beq
H_{j} = \left(
\begin{array}{cc}
\Omega_{j} & \mathsf{g}\\
\mathsf{g} & \varepsilon
\end{array}
\right).
\label{M}
\eeq
The eigenvalues of the matrix $H_{j}$ and the corresponding eigenvectors in the two dimensional subspace
$\{|\mathbf{G}\rangle \otimes |\tilde{1}_{j}\rangle, \,
|\vartheta_{j}\rangle \otimes |\mathbf{0}\rangle\}$ read, respectively, as
\beq
\Omega_{j, \,\pm} = \frac{\Omega{j} + \varepsilon}{2} \pm \chi_{j}, \;\;
\chi_{j} = \left(\left(\frac{\Delta_{j}}{2}\right)^{2} + \mathsf{g}^{2}\right)^{\frac{1}{2}},\;\;
\Delta_{j} = \varepsilon - \Omega_{j} = \delta - 2 \kappa \cos \Big(j\,\frac{2 \pi}{N}\Big),\;\;
\delta = \varepsilon - \Omega,
\label{eigenvalue}
\eeq
\beq
|\Omega_{j, \,\pm}\rangle =
\left(\left(\Omega_{j, \,\pm} - \varepsilon\right)^{2} + \mathsf{g}^{2}\right)^{-\frac{1}{2}}\;
\left(\left(\Omega_{j, \, \pm} - \varepsilon\right)\;
|\mathbf{G}\rangle \otimes |\tilde{1}_{j}\rangle
+ \mathsf{g}\,|\vartheta_{j}\rangle \otimes |\mathbf{0}\rangle\ \right).
\label{eigenstate}
\eeq
The detuning parameter $\delta$ measures the difference of the atomic frequency and that of the cavity photon,
whereas the corresponding differences for the delocalized modes are given by the parameters $\{\Delta_{j}| j = 0,1, \ldots, N-1\}$.
The time evolution of the coefficients in the expansion (\ref{state_deloc}) of the state $|\Psi (t)\rangle$
in the delocalized basis is given by
\beq
\left(\begin{array}{c}
\alpha_{j}(t)\\
\beta_{j}(t)
\end{array}\right)
= \exp\left(- i H_{j} t\right)
\left(\begin{array}{c}
\alpha_{j}(0)\\
\beta_{j}(0)
\end{array}\right)
\label{coeff_t},
\eeq
where the explicit construction reads
\bea
\alpha_{j}(t) &=& \exp\left(- i \frac{\Omega_{j} + \varepsilon}{2} \,t\right)\;
\sl{\left(\hbox{cos}\big(\chi_{j} t\big) + i\,\left(\frac{\Delta_{j}}{2}\right)\,
\frac{\hbox{sin}\big(\chi_{j} t\big)}{\chi_{j}}\right)\,\alpha_{j}(0)
- i\, \mathsf{g}\,\frac{\hbox{sin}\big(\chi_{j} t\big)}{\chi_{j}}\; \beta_{j}(0)},\nn\\
\beta_{j}(t) &=& \exp\left(- i \frac{\Omega_{j} + \varepsilon}{2} \,t\right)\;
\sl{\left(\hbox{cos}\big(\chi_{j} t\big) - i\,\left(\frac{\Delta_{j}}{2}\right)\,
\frac{\hbox{sin}\big(\chi_{j} t\big)}{\chi_{j}}\right)\,\beta_{j}(0)
- i\, \mathsf{g}\,\frac{\hbox{sin}\big(\chi_{j} t\big)}{\chi_{j}}\; \alpha_{j}(0)}.
\label{alpha_beta}
\eea

\par

To derive the effective Hamiltonian that discards the rapidly oscillating high-frequency
components on account of a time-averaging process, we use the interaction-picture scenario.
The interaction Hamiltonian given in (\ref{H0_decom}) now assumes the form:
\beq
{\cal H}^{\hbox{int}}(t) = \exp(i H_{0} t)\; H^{\hbox{int}}\,\exp(-i H_{0} t).
\label{H_int}
\eeq
Its explicit evaluation following from (\ref{H0_decom}-\ref{H_SA}) reads
\beq
{\cal H}^{\hbox{int}}(t) = \mathsf{g}\;\sum_{j = 0}^{N-1}
\left(\hbox{exp}\big(i\, \Delta_{j}\,  t\big)\; S_{j}^{+} A_{j}
+ \hbox{exp}\big(- i\, \Delta_{j} \,t\big)\;S_{j}^{-} A_{j}^{\dagger}\right).
\label{H_Ipic}
\eeq
As we are interested in dynamical processes occurring at low frequencies, all high frequency
components of the Hamiltonian are assumed to average out to zero resulting in a `coarse-grained'
effective Hamiltonian. Following the recipe given in
~\cite{JJ2007, OIK2008} the time-averaged effective Hamiltonian up to the
order $O(\mathsf{g}^{2})$
reads:
\beq
{\cal H}^{\hbox{eff}} = H_{0} + \left\langle{\cal H}^{\hbox{int}}(t)\right\rangle
+ \frac{1}{2}\, \left\langle\Big[{\cal H}^{\hbox{int}}(t), {\cal V}(t)\Big]\right\rangle
- \frac{1}{2}\,\Big[\Big\langle{\cal H}^{\hbox{int}}(t)\Big\rangle, \left\langle{\cal V}(t)\right\rangle\Big],
\label{H_g2}
\eeq
where the averaged value of a dynamical quantity $\mathcal{O}(t)$ with a suitable probability density function
$f(t)$ is defined ~\cite{JJ2007, OIK2008} as $\langle \mathcal{O}(t) \rangle =
\int_{-\infty}^{\infty} f(t-t^{\prime}) \mathcal{O}(t^{\prime}) dt^{\prime}$.
The generating element ${\cal V}(t)$ of the order $O(\mathsf{g}^{2})$ contributions in the rhs of (\ref{H_g2}) may be obtained
{\it \`{a} la} ~\cite{JJ2007, OIK2008}:
\beq
{\cal V}(t)= \mathsf{g}\;\sum_{j = 0}^{N-1}\;\frac{1}{\Delta_{j}}\;
\left(\hbox{exp}\big(- i\, \Delta_{j} \,t\big)\; S_{j}^{-} A_{j}^{\dagger}
- \hbox{exp}\big(i\, \Delta_{j}\, t\big)\;S_{j}^{+} A_{j}\right).
\label{V_Ipic}
\eeq
We now study various limiting cases of the effective Hamiltonian (\ref{H_g2}).

\par

The large hopping limit is dominated by the coupling between the photons of adjacent cavities:
$\kappa \gg \delta, \mathsf{g}$. We also assume $\Delta_{j} \gg \mathsf{g}\;\; \forall j \in \{0, 1, \ldots, N-1\}$.
These limits ensure vanishing of the following time-averaged quantities:
$ \left\langle{\cal H}^{\hbox{int}}(t)\right\rangle = 0,\;\; \left\langle{\cal V}(t)\right\rangle = 0$. Eliminating the
high frequency components via the averaging process the commutator  may be constructed using the definitions
(\ref{H_Ipic}) and (\ref{V_Ipic}):
\beq
\Big\langle\left[{\cal H}^{\hbox{int}}(t), {\cal V}(t)\right]\Big\rangle = 2\,\mathsf{g}^{2}\,
\Bigg(\sum_{j = 0}^{N - 1} \frac{1}{\Delta_{j}}\;\left[S_{j}^{+} A_{j}, S_{j}^{-} A_{j}^{\dagger}\right] +
\sum_{\atop{N > j > 0} {j \neq N-j}}\,\frac{1}{\Delta_{j}}\;
\left[S_{j}^{+} A_{j}, S_{N-j}^{-} A_{N-j}^{\dagger}\right]\Bigg).
\label{HV_com}
\eeq
The second term in the rhs of (\ref{HV_com}) owes its origin to the degeneracy of the eigenfrequencies
that is present for $N > 2$, and has been referred to in the context of (\ref{H0}). Explicit evaluation of the commutators following from (\ref{A_deloc}) and (\ref{S_com})
\bea
&&[S_{j}^{+} A_{j}, S_{j}^{-} A_{j}^{\dagger}] = \frac{1}{2}\,\left(S_{0}^{z}\,\left(A_{j}^{\dagger} A_{j} + A_{j} A_{j}^{\dagger}\right)
+ \left(S_{j}^{+} S_{j}^{-} + S_{j}^{-} S_{j}^{+}\right) \mathbb{I}\right),\nn\\
&&[S_{j}^{+} A_{j}, S_{N-j}^{-} A_{N-j}^{\dagger}] = S_{2j-N}^{z}\, A_{j} A_{N-j}^{\dagger}\qquad j \neq N-j \,\mod N
\label{comm_1}
\eea
allows us to obtain the effective Hamiltonian up to the order $O(\mathsf{g}^{2})$:
\bea
{\cal H}^{\hbox{eff}}  &=& H_{0} + \frac{\mathsf{g}^{2}}{2}\;\sum_{j = 0}^{N-1}
\frac{1}{\Delta_{j}} \,\left(S_{0}^{z}\,\left(A_{j}^{\dagger} A_{j} + A_{j} A_{j}^{\dagger}\right)
+ \left(S_{j}^{+} S_{j}^{-} + S_{j}^{-} S_{j}^{+}\right) \mathbb{I}\right)\nn\\
& &+ \mathsf{g}^{2} \sum_{\frac{N}{2} > j > 0} \frac{1}{\Delta_{j}}
\left(S_{2j-N}^{z}\, A_{j} A_{N-j}^{\dagger} + S_{N - 2 j}^{z}\, A_{j}^{\dagger} A_{N-j}\right).
\label{H_K_large}
\eea
The first term in the rhs of (\ref{H_K_large}) that may be read from (\ref{H0_decom}) describes the energy of the bare
delocalized atomic and field modes. The first sum represents the contribution due the Stark shift, and an induced
interaction of order $O(\mathsf{g}^{2})$ operating directly between the delocalized spin modes without the intermediacy of the photonic
modes. It is interesting to note that the delocalized modes maintain
the diagonal structure of the  spin-spin interaction term in the effective Hamiltonian.
The second sum is present due to the degeneracy of the field modes. The expression (\ref{H_K_large}) of the Hamiltonian
via the delocalized modes makes it evident that there is no transfer of energy
between the atomic and the photonic modes. Employing the reduction of the `averaged' spin coordinate
$S_{0}^{z}$ on the projectors $S_{0}^{z} = \frac{1}{N}\left(P_{e} - P_{g}\right)$, where
$P_{x} = \sum_{j = 0}^{N-1}|x_{j}> <x_{j}|\, \forall x\in \{g, e\}$, and with a suitable rearrangement of
the terms in the first sum in (\ref{H_K_large}) we recast  ${\cal H}^{\hbox{eff}}$ as follows:
\bea
{\cal H}^{\hbox{eff}}  &=& H_{0} + \frac{\mathsf{g}^{2}}{N}\;\s{P_{e}\;\sum_{j = 0}^{N-1}
\frac{1}{\Delta_{j}}\,A_{j} A_{j}^{\dagger} - P_{g}\;\sum_{j = 0}^{N-1}\frac{1}{\Delta_{j}}\,
A_{j}^{\dagger} A_{j} + \sum_{j = 0}^{N-1} \frac{1}{\Delta_{j}}\,\sum_{\atop{k, \ell = 0} {k \neq \ell}}^{N -1}
\omega^{- j (k - \ell)}\, \sigma_{k}^{+} \sigma_{\ell}^{-}\, \mathbb{I} \nn\\
& &+ \sum_{\frac{N}{2} > j > 0} \frac{1}{\Delta_{j}} \sum_{k = 0}^{N-1} \sigma_{k}^{z}
\Big(\omega^{-2 j k} \,A_{j} A_{N-j}^{\dagger} + \omega^{2 j k}\, A_{j}^{\dagger} A_{N-j}\Big)}.
\label{H_eff_spin}
\eea
In the above expression the Stark shift depends upon the population of the delocalized photonic modes.
The coupling of spins between any two arbitrary distinct site exist at the order $O(\mathsf{g}^{2})$. For
$N = 2$ the effective Hamiltonian (\ref{H_eff_spin}) agrees with the result obtained in \cite{OIK2008} at
the corresponding limit. To make the process transparent we introduce a discrete Fourier transform on the
reciprocal of the detuning parameters of the delocalized photonic modes:
\beq
G(j) = \frac{1}{N}\;\sum_{k = 0}^{N - 1}\frac{1}{\Delta_{k}}\;\omega^{- j k},
\label{Green}
\eeq
and thereby reexpress (\ref{H_eff_spin}) in the following form:
\bea
{\cal H}^{\hbox{eff}}  &=& H_{0} + \frac{\mathsf{g}^{2}}{N}\;\s{P_{e}\;\sum_{j, k = 0}^{N-1}
a_{j}\overline{G(j - k)}\,a_{k}^{\dagger} - P_{g}\;\sum_{j, k = 0}^{N-1}
a_{j}^{\dagger}\, G(j - k)\,a_{k} + N \;\sum_{\atop{j, k = 0}{j \neq k}}^{N-1}\sigma_{j}^{+}\, G(j - k)\, \sigma_{k}^{-}\;\mathbb{I}\nn\\
& &  + \sum_{\frac{N}{2} > j > 0} \frac{1}{\Delta_{j}} \sum_{k = 0}^{N-1} \sigma_{k}^{z}
\Big(\omega^{-2 j k} \,A_{j} A_{N-j}^{\dagger} + \omega^{2 j k}\, A_{j}^{\dagger} A_{N-j}\Big)}.
\label{HK_Green}
\eea
It is evident from the structure of (\ref{HK_Green}) that the Fourier transform $G(j)$ acts as the propagator
of the atomic and the photonic excitations. The transition amplitudes of these excitations proceeding on the loop are
proportional to the magnitude of the propagator.

\par

To further examine the issue of transfer of the excitation modes we turn to the solutions (\ref{alpha_beta}). The following
limits valid in the present approximation
\beq
\frac{\Delta_{j}}{2 \chi_{j}}\Big|_{\mathsf{g} \ll \frac{|\Delta_{j}|}{2}} \longrightarrow
\frac{\Delta_{j}}{|\Delta_{j}|},\qquad
\frac{\mathsf{g}}{\chi_{j}}\Big|_{\mathsf{g} \ll \frac{|\Delta_{j}|}{2}} \longrightarrow 0,
\label{Del_lim}
\eeq
yields the structure
\beq
\alpha_{j}(t) = \exp\left(- i \left(\Omega_{j} - \frac{\mathsf{g}^{2}}{\Delta_{j}}\right) t\right)\;\alpha_{j}(0),\qquad
\beta_{j}(t) = \exp\left(- i \left(\varepsilon + \frac{\mathsf{g}^{2}}{\Delta_{j}}\right)\,t\right)\;\beta_{j}(0).
\label{Al_Be_lim}
\eeq
It follows from the above equations for the delocalized modes that the atomic and photonic excitations remain
decoupled from each other.
Employing these solutions with the Fourier expansions (\ref{albe_exp}, \ref{ab_exp}) we, in this limit,
obtain the time-evolution equation for the coefficients of the localized single excitation states:
\bea
\mathsf{a}_{j}(t) &=& \frac{1}{N} \sum_{k, \ell = 0}^{N-1}\;
\exp\s{-i \left(\frac{2 \pi}{N}\,(j - k) \ell + \left(\Omega_{\ell} - \frac{{g}^{2}}{{\Delta}_{\ell}}\right)\, t\right)}\, \mathsf{a}_{k}(0),
\nn\\
\mathsf{b}_{j}(t) &=& \frac{1}{N} \sum_{k, \ell = 0}^{N-1}\;
\exp\s{-i \left(\frac{2 \pi}{N}\,(j - k) \ell + \left(\varepsilon + \frac{{g}^{2}}{{\Delta}_{\ell}}\right)\, t\right)}\, \mathsf{b}_{k}(0).
\label{AB_evol}
\eea
Equation (\ref{AB_evol}) indicates that subject to the approximation (\ref{Del_lim}) transfer of excitations from the atomic to
the photonic modes or vice versa does not take place, and, consequently, the probability density in each kind of degree of freedom is conserved:
\beq
\sum_{j = 0}^{N-1}\,|\mathsf{a}_{j}(t)|^{2} =  \sum_{j = 0}^{N-1}\,|\mathsf{a}_{j}(0)|^{2},\qquad
\sum_{j = 0}^{N-1}\,|\mathsf{b}_{j}(t)|^{2} =  \sum_{j = 0}^{N-1}\,|\mathsf{b}_{j}(0)|^{2}.
\label{ab_conserv}
\eeq
For definiteness, we assume that at $t = 0$ only the $0$-th atom is in the excited state and all other
degrees of freedom are in the ground state:   $\mathsf{a}_{j}(0) = 0,\;
\mathsf{b}_{j}(0) = \delta_{j 0}$. Evolution equations in (\ref{AB_evol}) now reduce to
\beq
\mathsf{a}_{j}(t) = 0,\qquad \mathsf{b}_{j}(t) = \frac{1}{N} \sum_{k = 0}^{N-1}\;
\exp\s{-i \left(j k \frac{2 \pi}{N} + \left(\varepsilon + \frac{{g}^{2}}{{\Delta}_{k}}\right)\, t\right)}.
\label{b_exp}
\eeq
For $N \neq 2$ exact transmission of the excitation does not take place.

\par

We next study the large detuning limit: $\delta \gg \kappa, \mathsf{g}$ when the atoms are
highly detuned from the photonic modes. In this limit also the following time-averaged properties hold:
$\left\langle{\cal H}^{\hbox{int}}(t)\right\rangle = 0,\;\; \left\langle{\cal V}(t)\right\rangle = 0$.
Towards obtaining the effective Hamiltonian we first evaluate the time-averaged commutator in this limit:
\beq
\Big\langle\left[{\cal H}^{\hbox{int}}(t), {\cal V}(t)\right]\Big\rangle = \mathsf{g}^{2}\,
\sB{2\,\sum_{j = 0}^{N - 1} \frac{1}{\Delta_{j}}\;\left[S_{j}^{+} A_{j}, S_{j}^{-} A_{j}^{\dagger}\right] +
\sum_{\atop{ j, k = 0} {j \neq k}}^{N - 1}\,\left(\frac{1}{\Delta_{j}} + \frac{1}{\Delta_{k}}\right)\;
\left[S_{j}^{+} A_{j}, S_{k}^{-} A_{k}^{\dagger}\right]}.
\label{HV_com _2}
\eeq
Employing the first equation in (\ref{comm_1}) and rearranging terms on the rhs of (\ref{HV_com _2}) the said
commutator is reexpressed as
\beq
\Big\langle\left[{\cal H}^{\hbox{int}}(t), {\cal V}(t)\right]\Big\rangle = \mathsf{g}^{2}\,\sB{
\sum_{j, k = 0}^{N - 1}\,\frac{1}{\Delta_{j}} \,\left(S_{j - k}^{z}\,A_{j} A_{k}^{\dagger} + S_{k - j}^{z}\, A_{j}^{\dagger} A_{k}\right)
+ \sum_{j = 0}^{N - 1}\,\frac{1}{\Delta_{j}} \,\left(S_{j}^{+} S_{j}^{-} + S_{j}^{-} S_{j}^{+}\right) \mathbb{I}}.
\label{HV_D_large}
\eeq
The effective Hamiltonian up to the order $O(\mathsf{g}^{2})$ may now be obtained via (\ref{H_g2}) and (\ref{HV_D_large}):
\beq
{\cal H}^{\hbox{eff}}  = H_{0} + \frac{\mathsf{g}^{2}}{2}\,
\sB{\sum_{j, k = 0}^{N - 1}\,\frac{1}{\Delta_{j}} \,\left(S_{j - k}^{z}\,A_{j} A_{k}^{\dagger}
+ S_{k - j}^{z}\, A_{j}^{\dagger} A_{k}\right)
+ \sum_{j = 0}^{N - 1}\,\frac{1}{\Delta_{j}} \,\left(S_{j}^{+} S_{j}^{-} + S_{j}^{-} S_{j}^{+}\right) \mathbb{I}}.
\label{H_D_large}
\eeq
In the $N = 2$ case the above expression reduces to the results obtained in \cite{OIK2008} in the large detuning limit.
The effective Hamiltonian (\ref{H_D_large}) may be recast using the propagator in the Fourier space (\ref{Green}), and the transforms
(\ref{A_deloc}, \ref{S_deloc}):
\bea
{\cal H}^{\hbox{eff}}  &=& H_{0} + \frac{\mathsf{g}^{2}}{2}\,\sum_{j, k = 0}^{N - 1} \s{\sigma_{k}^{z}\;
\left(a_{j}^{\dagger}\,G(j - k)\,a_{k} + a_{j}\overline{G(j - k)}\,a_{k}^{\dagger}\right)\nn\\
& &+ \left(\sigma_{j}^{+}\,G(j - k)\,\sigma_{k}^{-}
+ \sigma_{j}^{-}\,\overline{G(j - k)}\,\sigma_{k}^{+}\right)\;\mathbb{I}}.
\label{HG-loc}
\eea
In this approximation also the atomic and the photonic modes remain decoupled from each other forbidding any transfer of energy
up to the order $O(\mathsf{g}^{2})$. The time-evolution of the coefficients $\mathsf{a}_{j}(t), \mathsf{b}_{j}(t)$ of the
localized single-excitation states reduces to previously described form (\ref{AB_evol}). Assuming dominant value of the detuning parameter
$\delta \gg \kappa, \mathsf{g}$ we now study the asymptotic  $N \rightarrow \infty$
limit of (\ref{AB_evol}) for the boundary condition $\mathsf{a}_{j}(0) = \delta_{j 0}, \mathsf{b}_{j} (0)= 0$. Using the Fourier
transform
\beq
\exp (\pm i x\, \cos \theta) = \sum_{n = - \infty}^{\infty}\, (\pm i)^{n}\; J_{n}(x)\; \exp(\pm i n \theta),\qquad
J_{n}(x) = (-1)^{n} \;J_{n}(x)
\label{Fourier_B}
\eeq
via the Bessel functions $J_{n}(x)$, and employing the said boundary condition the time evolution of the coefficients of the
localized states may be expressed as
\beq
\mathsf{a}_{j} (t) = \exp \s{- i \Big(\Omega - \frac{\mathsf{g}^{2}}{\delta}\Big) t} \,
\sum_{\nu = - \infty}^{\infty} (- i)^{j + \nu N}\; J_{j + \nu N}\s{2 \kappa \Big(1 - \frac{\mathsf{g}^{2}}{\delta^{2}}\Big) t},\quad
\mathsf{b}_{j} (t) = 0.
\label{del_large}
\eeq
In deriving (\ref{del_large}) we have used the summation (\ref{unitary}). The asymptotic expansion of the Bessel function
of large order \cite{AS1964} $J_{N}(N)|_{N \rightarrow \infty} \sim N ^{- 1/3}$ allows us to obtain an asymptotic scaling limit of the
magnitude of the transition coefficient $|\mathsf{a}_{N/2} (t \sim N)|_{N \rightarrow \infty} \sim N ^{- 1/3}$.

\par

The case of comparable hopping and detuning is of particular interest. Turning towards this near-coherence scenario
where the atomic frequency approximately equals to  the eigenfrequncy  of,
say the $\ell$-th delocalized photonic
mode: $\varepsilon \approx \Omega_{\ell}\;\Rightarrow \Delta_{\ell} \approx 0$, we observe that the degeneracy (\ref{H0})
of the photonic eigenmodes present when $\ell \neq N - \ell\, \mod N$ needs to be taken into account.
We also assume that the detuning parameters for all the nonresonant modes are large:
$\Delta_{j} \gg 0 \;\,\forall j \notin \{\ell, N-\ell (\neq \ell \mod N)\}$. Unlike
the previous cases considered the present near-resonance limit allows a direct transfer of the excitation between the atomic and
the photonic modes. The time-averaged contribution of the near-resonance mode to the ${\cal H}^{\hbox{int}}$ at the order
$O({\mathsf g})$ may be read from (\ref{H_Ipic}):
\beq
\left\langle{\cal H}^{\hbox{int}}(t)\right\rangle = \mathsf{g}\,\sq{\big(S_{\ell}^{+} A_{\ell}
+ S_{\ell}^{-} A_{\ell}^{\dagger}\big) + \big(1 - \delta_{\ell, N - \ell\; \hbox{mod}\, N}\big)\;
\big(S_{N - \ell}^{+} A_{N - \ell}+ S_{N - \ell}^{-} A_{N - \ell}^{\dagger}\big)},
\label{H_reso}
\eeq
whereas its contribution to  $\left\langle{\cal V}(t)\right\rangle$ follows from (\ref{V_Ipic}):
\beq
\left\langle{\cal V}(t)\right\rangle = \frac{\mathsf{g}}{\Delta_{\ell}}
\sq{\big(S_{\ell}^{-} A_{\ell}^{\dagger} - S_{\ell}^{+} A_{\ell}\big)
+ \big(1 - \delta_{\ell, N - \ell\;\hbox{mod}\, N}\big)\;
\big(S_{N - \ell}^{-} A_{N - \ell}^{\dagger} - S_{N - \ell}^{+} A_{N - \ell}\big)}.
\label{V_reso}
\eeq
When a near-resonance scenario involving these degenerate eigenmodes are realized the said modes give rise to independent
contributions associated with them, respectively. The commutator of the above time-averaged quantities reads
\bea
\left[\left\langle{\cal H}^{\hbox{int}}(t)\right\rangle, \left\langle{\cal V}(t)\right\rangle\right]
&=& \frac{2 \mathsf{g}^{2}}{\Delta_{\ell}}\;\left(\left[S_{\ell}^{+} A_{\ell}, S_{\ell}^{-} A_{\ell}^{\dagger}\right]
+ (1-\delta_{\ell, N - \ell\; \hbox{mod}\, N})\,\Lambda_{N, \ell}\right),\nn\\
\Lambda_{N, \ell} &=& \left[S_{N - \ell}^{+} A_{N - \ell}, S_{N - \ell}^{-} A_{N - \ell}^{\dagger}\right]
+ S_{2 \ell - N}^{z}\,A_{\ell} A_{N - \ell}^{\dagger} + S_{N - 2 \ell}^{z}\,A_{\ell}^{\dagger} A_{N - \ell}.
\label{HV3_comm}
\eea
Combining the previous results the effective Hamiltonian (\ref{H_g2}) in the near-resonance case up to the order
$O(\mathsf{g}^{2})$  assumes the form
\bea
{\cal H}^{\hbox{eff}} &=& H_{0} + \mathsf{g}\,\sq{\big(S_{\ell}^{+} A_{\ell}
+ S_{\ell}^{-} A_{\ell}^{\dagger}\big) + \big(1 - \delta_{\ell, N - \ell\;\hbox{mod}\, N}\big)\;
\big(S_{N - \ell}^{+} A_{N - \ell}+ S_{N - \ell}^{-} A_{N - \ell}^{\dagger}\big)}\nn\\
& & + \frac{\mathsf{g}^{2}}{2}\;\sideset{}{'}\sum_{j = 0}^{N-1}
\frac{1}{\Delta_{j}} \,\sq{S_{0}^{z}\,\left(A_{j}^{\dagger} A_{j} + A_{j} A_{j}^{\dagger}\right)
+ \left(S_{j}^{+} S_{j}^{-} + S_{j}^{-} S_{j}^{+}\right) \mathbb{I}}\nn\\
& &+ \mathsf{g}^{2} \;\sideset{}{''}\sum_{\frac{N}{2} > j > 0} \frac{1}{\Delta_{j}}
\sq{S_{2j-N}^{z}\, A_{j} A_{N-j}^{\dagger} + S_{N - 2 j}^{z}\, A_{j}^{\dagger} A_{N-j}},
\label{H_DKeq}
\eea
where the first sum on the rhs excludes the resonance modes $\{\ell, (N - \ell)\;\hbox{when}\; \ell \neq (N-\ell) \mod N\}$,
and  the resonance mode ($\ell$) lying in the domain of the index of the second sum is eliminated.
The term $O(\mathsf{g})$ in this Hamiltonian describes a transfer of energy between the near-resonant delocalized
photon modes and their spin excitation partners. The non-resonant modes contributing in the first sum at order $O(\mathsf{g}^{2})$
represent the Stark shift of the atoms and a direct transfer of excitations between the atomic modes without the
intermediacy  of the field modes. In the near-resonance condition we assume the detuning parameter of the resonant modes are negligible
compared to the atom-cavity photon coupling:
$\Big|\frac{\Delta_{\ell}}{\mathsf{g}}\Big| \ll 1  \; \forall j \in \{\ell, (N - \ell) \;\hbox{when}\;\ell \neq (N-\ell) \mod N\}$. Excitations of the
near-resonant field modes and that of their coupled spin partners may be read from (\ref{alpha_beta}):
\bea
\alpha_{j}(t) &=& \exp\left(- i \left(\varepsilon -\frac{\Delta_{\ell}}{2}\right) t\right)\;
\sB{\cos \left(\left(\mathsf{g} + \frac{\Delta_{\ell}^{2}}{8 \mathsf{g}}\right)\,t \right)\;
\alpha_{j}(0) - i \,\sin \left(\left(\mathsf{g} + \frac{\Delta_{\ell}^{2}}{8 \mathsf{g}}\right)\,t \right)\;
\beta_{j}(0)},\nn\\
\beta_{j}(t) &=& \exp\left(- i \left(\varepsilon -\frac{\Delta_{\ell}}{2}\right) t\right)\;
\sB{\cos \left(\left(\mathsf{g} + \frac{\Delta_{\ell}^{2}}{8 \mathsf{g}}\right)\,t \right)\; \beta_{j}(0)
- i \, \sin \left(\left(\mathsf{g} + \frac{\Delta_{\ell}^{2}}{8 \mathsf{g}}\right)\,t \right)\;
\alpha_{j}(0)}.
\label{AlBe_res}
\eea
Assuming that the detuning parameters for the non-resonant modes satisfy $\Big|\frac{\mathsf{g}}{\Delta_{j}}\Big| \ll 1$, where
$j \notin \{\ell, (N - \ell) \;\hbox{when}\;\ell \neq (N-\ell) \mod N\}$, the corresponding excitations assume the form (\ref{Al_Be_lim}).

\par

The above evolution equations for the delocalized modes in conjunction with the Fourier transforms (\ref{albe_exp}, \ref{ab_exp})
now yield the single-excitation states of the atoms and the cavity photons. We first assume that the
near resonant $\ell$-th mode is nondegenerate: $\ell = (N - \ell)\;\hbox{mod}\, N$. The $\ell = 0, N/2 \;(\hbox{for even}\, N)$\, modes
satisfy this property. For the choice of the initial excitations being atomic in nature: $\mathsf{a}_{j}(0) = 0$, the time evolution of
the coefficients are given by
\bea
\mathsf{a}_{j}(t) &=& - \frac {i}{N}\,\exp\sl{-i\left(\varepsilon - \frac{\Delta_{\ell}}{2}\right) t}\;
\sin\sl{\left(\mathsf{g} +\frac{\Delta_{\ell}^{2}}{8 \mathsf{g}}\right) t} \;
\sum_{k = 0}^{N - 1} \omega^{-\ell(j - k)}\,\mathsf{b}_{k}(0),\nn\\
\mathsf{b}_{j}(t) &=& \frac{1}{N} \sum_{k = 0}^{N - 1}\,\s{\sum_{\atop{n = 0}{n \neq \ell}}^{N - 1}\,
\omega^{- n(j -k)}\,\exp\left(-i\left(\varepsilon + \frac{\mathsf{g}^{2}}{\Delta_{n}}\right) t\right)\nn\\
& & +  \omega^{-\ell(j - k)} \;\exp\left(-i\left(\varepsilon - \frac{\Delta_{\ell}}{2}\right) t\right)\;
\cos\left(\left(\mathsf{g} +\frac{\Delta_{\ell}^{2}}{8 \mathsf{g}}\right) t\right)}\;\mathsf{b}_{k}(0).
\label{off_res-1}
\eea
In the exact resonance condition $\Delta_{\ell} = 0$, and in a scenario where the dispersive effects are negligible
$\Big|\frac{\mathsf{g}}{\Delta_{j}}\Big| \approx 0 \;\;\forall\, j \neq \ell$  the above solutions assume the form
\bea
\mathsf{a}_{j}(t) &=& - \frac{i}{N}\,\exp(-i \varepsilon t)\,\sin(\mathsf{g} t)\;
\sum_{k = 0}^{N - 1} \omega^{-\ell(j - k)}\,\mathsf{b}_{k}(0),\nn\\
\mathsf{b}_{j}(t) &=& \exp(-i \varepsilon t)\,\sl{\mathsf{b}_{j}(0) - \frac{2}{N}\,
\sin^{2}\left(\frac{\mathsf{g} t}{2}\right)\,\sum_{k = 0}^{N - 1} \omega^{-\ell(j - k)}
\,\mathsf{b}_{k}(0)}.
\label{ab_res}
\eea
 With the further assumption
that only the $0$-th atom is initially excited whereas all other atoms remain in the ground state
$\mathsf{b}_{j}(0) = \delta_{j 0}\; \forall j \in \{0, 1, \ldots, N-1\}$, the coefficients of the localized
single-excitation states read
\bea
\mathsf{a}_{j}(t) &=& - \frac{i}{N}\,\sin(\mathsf{g} t)\,\exp\sl{-i \left(j \ell \frac{2 \pi}{N}
+ \varepsilon t\right)},\nn\\
\mathsf{b}_{j}(t) &=& \exp(-i \varepsilon t)\,\delta_{j 0} - \frac{2}{N}\;\sin^{2}\left(\frac{\mathsf{g} t}{2}\right)\;
\exp\sl{-i \left(j \ell \frac{2 \pi}{N} + \varepsilon t\right)}.
\label{ab_evol}
\eea
For $N = 2$ the magnitude of the localized atomic excitation $\mathsf{b}_{1}(t)$ assumes unit value for both choices of
the resonance modes $\ell = 0, 1$ at the specified time $t = (2 n + 1) \pi/\mathsf{g}\; \forall n \in \mathbb{Z}_{+}$,
making the transmission of the state exact.

\par

Turning towards the resonance of the atomic mode with a conjugate pair of degenerate delocalized photonic modes $\{\ell, (N - \ell)\}$,
where $\ell \neq (N - \ell) \mod N$, we, following the preceding recipe, construct the time-evolutions of the excitations of the atoms
and the cavity-photons:
\bea
\mathsf{a}_{j}(t) &=& - \frac{2 i}{N}\,\exp\sl{-i\left(\varepsilon - \frac{\Delta_{\ell}}{2}\right) t}\;
\sin\sl{\left(\mathsf{g} +\frac{\Delta_{\ell}^{2}}{8 \mathsf{g}}\right) t} \;
\sum_{k = 0}^{N - 1} \cos\left(\frac{2 \pi}{N}\,\ell(j - k)\right)\,\mathsf{b}_{k}(0),\nn\\
\mathsf{b}_{j}(t) &=& \frac{1}{N} \sum_{k = 0}^{N - 1}\,\s{\sum_{\atop{n = 0}{n \neq \ell, N - \ell}}^{N - 1}\,
\omega^{- n(j -k)}\,\exp\left(-i\left(\varepsilon + \frac{\mathsf{g}^{2}}{\Delta_{n}}\right) t\right)\nn\\
& & + 2\, \exp\left(-i\left(\varepsilon - \frac{\Delta_{\ell}}{2}\right) t\right)\;
\cos\left(\left(\mathsf{g} +\frac{\Delta_{\ell}^{2}}{8 \mathsf{g}}\right) t\right) \;
\cos\left(\frac{2 \pi}{N}\,\ell(j - k)\right)}\mathsf{b}_{k}(0).
\label{off_res-2}
\eea
In the above equation we have assumed that the initial excitations are only atomic in nature. For the exact resonance condition
$\Delta_{\ell} = 0$ and in the dispersion-free limit $\frac{\mathsf{g}}{\Delta_{j}} \approx 0 \;\;\forall\, j \notin \{\ell, (N - \ell)\}$
the excitations reduce to the form
\bea
\mathsf{a}_{j}(t) &=& - \frac{2 i}{N}\,\exp(-i \varepsilon t)\,\sin(\mathsf{g} t)\;
\sum_{k = 0}^{N - 1} \cos\left(\frac{2 \pi}{N}\,\ell(j - k)\right)\,\mathsf{b}_{k}(0),\nn\\
\mathsf{b}_{j}(t) &=& \exp(-i \varepsilon t)\,\sB{\mathsf{b}_{j}(0) - \frac{4}{N}\,
\sin^{2}\left(\frac{\mathsf{g} t}{2}\right)\;\sum_{k = 0}^{N - 1}\,
\cos\left(\frac{2 \pi}{N}\,\ell(j - k)\right)\,\mathsf{b}_{k}(0)}.
\label{abdeg_res}
\eea
Restraining to the case where only the $0$-th atom is initially excited
$\mathsf{b}_{j}(0) = \delta_{j 0}\;\forall j \in \{0, 1, \ldots, N-1\}$ the time evolution of the coefficients read
\bea
\mathsf{a}_{j}(t) &=& - \frac{i}{N}\,\sin(\mathsf{g} t)\,\s{\exp\left(i \left(j \ell \frac{2 \pi}{N}
- \varepsilon t\right)\right)+ \exp\left(-i \left(j \ell \frac{2 \pi}{N}
+ \varepsilon t\right)\right)},\nn\\
\mathsf{b}_{j}(t) &=& \exp(-i \varepsilon t)\,\delta_{j 0} - \frac{2}{N}\;\sin^{2}\left(\frac{\mathsf{g} t}{2}\right)\;
\s{\exp\left(i \left(j \ell \frac{2 \pi}{N} - \varepsilon t\right)\right)
+ \exp\left(-i \left(j \ell \frac{2 \pi}{N} + \varepsilon t\right)\right)},
\label{abdeg_evol}
\eea
where a  superposition of the clockwise and the anticlockwise modes takes place. For $N = 4$ and with the choice of the
resonance mode $\ell = 1$ the magnitude of the excitation $\mathsf{b}_{2}(t)$ assumes unit value at
$t = (2 n + 1) \pi/\mathsf{g}\;\; \forall n \in \mathbb{Z}_{+}$.

\section{Linear chain with parabolic coupling}
\label{parabolic}
\setcounter{equation}{0}
Interesting situations arise if one considers nearest-neighbor couplings between the cavities to be nonuniform in
nature. In particular it has been observed \cite{MCHGH2009} that a parabolic coupling between the photons of adjacent
cavities gives rise to dispersion free transmission of the excitations along one dimensional chain of cavities. The inter-cavity
photonic hopping term of the Hamiltonian now reads
\beq
H^{\hbox{hop}} = \kappa \sum_{j, k = 0}^{N-1} a_{j}^{\dagger}\, {\mathsf C}_{j, k} \,a_{k}, \qquad
{\mathsf C}_{j, k} = \sqrt{j (N - j)}\;\delta_{j\,k+1} + \sqrt{k (N - k)}\;\delta_{j+1\,k}.
\label{hop_K}
\eeq
The photonic part of the interaction Hamiltonian relates to the tridiagonal Jacobi matrix of the Krawtchouk polynomials \cite{KS1998}.
Therefore the photonic degrees of freedom may be diagonalized using delocalized wave functions expressed via these discrete
orthogonal polynomials. Very briefly we now introduce the standard notations on the  Krawtchouk polynomials. Detailed
discussions may be obtained from Ref. \cite{KS1998}.

\par

The Krawtchouk polynomial of degree $n$ ($n=0,1,\ldots,\mathfrak{N}$) in the variable $x$, with parameter $0<p<1$ is given by
\beq
K_n(x)\equiv K_n(x; p,\mathfrak{N}) = \mbox{$_2F_1$} \left(\atop{-x,-n}{-\mathfrak{N}} ; \frac{1}{p}\right).
\label{K_def}
\eeq
The function ${}_2F_1$ is the classical hypergeometric series~\cite{Bailey1964, Slater1966},
and in this case it is a terminating series because of the appearance of the negative integer $(-n)$ as a numerator parameter. It
is convenient to introduce orthonormal Krawtchouk polynomials by
\beq
\widetilde{K}_n(x) \equiv \frac{\sqrt{w(x)} K_n(x)}{\sqrt{d_n}},
\label{K-tilde}
\eeq
where $w(x)$ is the weight function in $x$, and $d_n$ is a function depending on $n$:
\beq
w(x) = \binom{\mathfrak{N}}{x} \, p^x \, (1-p)^{\mathfrak{N}-x}\quad \forall \;x=0,1,\ldots,\mathfrak{N}; \qquad\qquad
d_n = \frac{1}{\binom{\mathfrak{N}}{n}} \left( \frac{1-p}{p} \right)^n.
\label{wd}
\eeq
The scaled polynomials $\widetilde{K}_n(x)$ satisfy a discrete orthogonality relation~\cite{KS1998}:
\beq
\sum_{x=0}^{\mathfrak{N}} \widetilde{K}_n(x) \widetilde{K}_m(x) = \delta_{n m}.
\label{K_ortho}
\eeq

\par

In the present scenario of parabolic coupling of the photons of adjacent cavities the  Krawtchouk polynomials of parametric
value $p = 1/2$ play an essential role in diagonalizing the photonic part of the Hamiltonian. Following \cite{RV2009} we introduce
a symmetric orthogonal $N \times N$ matrix that is comprised of Krawtchouk polynomials as
\beq
\mathsf{U}_{jk} = \widetilde {K}_{k}\left(j;\frac{1}{2}; N-1\right)\;\; \forall\; j,k = 0, 1,\ldots, N-1,\qquad \mathsf{U} = \mathsf{U}^{T},
\qquad \mathsf{U} \mathsf{U}^{T} = \mathsf{U}^{T}\,\mathsf{U}=\mathbb{I}.
\label{KU_def}
\eeq
The adjacency matrix ${\mathsf C}$ given in (\ref{hop_K}) may now be diagonalized via the orthogonal matrix $\mathsf{U}$ as follows:
\beq
\mathsf{U} \,\mathsf{C}\,\mathsf{U}^{T} = D,\qquad\qquad D= \diag (N-1,\, N-3,\ldots, -(N-1)).
\label{KC_diag}
\eeq
We note that the parity symmetry of the scaled Krawtchouk polynomials described below
\beq
\widetilde {K}_{j}\left(\ell;\frac{1}{2}; \mathfrak{N}\right) = (-1)^{\ell}\;
\widetilde {K}_{\mathfrak{N} - j}\left(\ell;\frac{1}{2}; \mathfrak{N}\right),\quad
\widetilde {K}_{\ell}\left(j;\frac{1}{2}; \mathfrak{N}\right) = (-1)^{\ell}\;
\widetilde {K}_{\ell}\left(\mathfrak{N} - j;\frac{1}{2}; \mathfrak{N}\right)
\;\;\forall\; j \ne \mathfrak{N} - j
\label{KU-mirror}
\eeq
interrelates the components of the orthogonal matrix $\mathsf{U}$:
\beq
\mathsf{U}_{N - 1 \; j} = (-1)^{j}\; \mathsf{U}_{0 \; j}.
\label{KU_rel}
\eeq

\par

The delocalized collective modes may now be introduced as transform of the corresponding localized degrees of freedom via the
orthogonal symmetric matrix $\mathsf{U}$ as given below:
\bea
\mathsf{A}_{j} &=& \sum_{k = 0}^{N - 1} \mathsf{U}_{j k}\,a_{k}, \qquad
\mathsf{A}_{j}^{\dagger} = \sum_{k = 0}^{N - 1} \mathsf{U}_{j k}\,a_{k}^{\dagger} \qquad \Rightarrow
\qquad [\mathsf{A}_{j}, \mathsf{A}_{k}^{\dagger}] = \delta_{j k},\nn\\
\mathsf{S}_{j}^{+} &=& \sum_{k = 0}^{N - 1} \mathsf{U}_{j k}\,\sigma_{k}^{+}, \;\;\;
\mathsf{S}_{j}^{-} = \sum_{k = 0}^{N - 1} \mathsf{U}_{j k}\,\sigma_{k}^{-},\;\;\;
\mathsf{S}_{j k}^{z} = \sum_{\ell = 0}^{N - 1} \mathsf{U}_{j \ell}\,\mathsf{U}_{k \ell}\,\sigma_{\ell}^{z}\;\;
\Rightarrow\;\; \left[\mathsf{S}_{j}^{+}, \mathsf{S}_{k}^{-}\right] = \mathsf{S}_{j k}^{z}.
\label{comm_2}
\eea
The photonic degrees of freedom $\{\mathsf{A}_{j}| j \in 0,1, \ldots N-1\}$ satisfy the Heisenberg algebra. We now
recast the Hamiltonian employing  these delocalized variables. The rotating wave structure of its
atom-photon interaction term is maintained in terms of the delocalized modes:
\beq
H_{0} = \varepsilon\,\sum_{j=0}^{N-1} \mathsf{S}_{j}^{+}\,\mathsf{S}_{j}^{-}
+ \sum_{j=0}^{N-1} \widehat{\Omega}_{j} \mathsf{A}_{j}^{\dagger}\,\mathsf{A}_{j}, \qquad
H^{\hbox{int}} = \mathsf{g}\,\sum_{j=0}^{N-1} \left(\mathsf{S}_{j}^{+}\,\mathsf{A}_{j}
+ \mathsf{S}_{j}^{-}\,\mathsf{A}_{j}^{\dagger}\right),
\label{K_H}
\eeq
where the spectrum of the  delocalized photonic  eigenmodes is given by
$\widehat{\Omega}_{j} = \Omega \,+ \,\kappa \,(N - 1 -2 j)$ for $j = 0, 1, \ldots, N-1$.
The delocalized single excitation atomic and photonic states are constructed by the action of the collective
operators introduced in (\ref{comm_2}):
\bea
|\hat{1}_{j}\rangle &=& \mathsf{A}_{j}^{\dagger} |\mathbf{0}\rangle =
\sum_{k = 0}^{N - 1} \mathsf{U}_{j k}\,|1_{k}\rangle, \qquad
\mathsf{A}_{j} |\hat{1}_{k}\rangle = \delta_{j k} |\mathbf{0}\rangle, \qquad
\langle \hat{1}_{j}|\hat{1}_{k}\rangle = \delta_{j k}, \qquad
\langle 1_{j}|\hat{1}_{k}\rangle = \mathsf{U}_{j k},\nn\\
|\hat{\vartheta}_{j}\rangle &=& \mathsf{S}_{j}^{+} |\mathbf{G}\rangle =
\sum_{k = 0}^{N - 1} \mathsf{U}_{j k}\,|e_{k}\rangle, \qquad
\mathsf{S}_{j}^{-} |\hat{\vartheta}_{k}\rangle = \delta_{j k} |\mathbf{G}\rangle, \qquad
\langle \hat{\vartheta}_{j}|\hat{\vartheta}_{k}\rangle = \delta_{j k}, \qquad
\langle e_{j}|\hat{\vartheta}_{k}\rangle = \mathsf{U}_{j k}.
\label{K_state}
\eea
In the above delocalized basis an arbitrary single excitation state may be expanded as follows:
\beq
|\Psi (t)\rangle = \sum_{j = 0}^{N-1}\;
\left(\hat{\alpha}_{j}(t)\;|\mathbf{G}\rangle \otimes |\hat{1}_{j}\rangle \;
+ \;\hat{\beta}_{j}(t)\;|\hat{\vartheta}_{j}\rangle \otimes |\mathbf{0}\rangle\right).
\label{Kstate_deloc}
\eeq
Invertible transformation via the orthogonal symmetric matrix $\mathsf{U}$ interrelate the coefficients
of the above expansions  with the the corresponding coefficients of the expansion (\ref{state_loc}) in the basis of the
localized single excitation states:
\beq
\hat{\alpha}_{j}(t) = \sum_{k = 0}^{N-1}\; \mathsf{U}_{j k}
\mathsf{a}_{k}(t),\quad
\hat{\beta}_{j}(t) = \sum_{k = 0}^{N-1}\; \mathsf{U}_{j k}
\mathsf{b}_{k}(t),\quad
\mathsf{a}_{j}(t) = \sum_{k = 0}^{N-1}\; \mathsf{U}_{j k}
\hat{\alpha}_{k}(t),\quad
\mathsf{b}_{j}(t) = \sum_{k = 0}^{N-1}\;\mathsf{U}_{j k}
\hat{\beta}_{k}(t).
\label{Kab_exp}
\eeq

\par

We first consider the regime of dominant hopping parameter: $\kappa\gg \delta$, whereas the atom-photon coupling
is assumed to be small: $\Big|\frac{\mathsf{g}}{\hat{\Delta}_{j}}\Big| \ll 1 \;\;\forall j \in (0, 1, \ldots, N-1)$. Unlike
the case considered in Sec. \ref{cyclic} the parabolic coupling between the photons of adjacent cavities does not
produce degenerate eigenvalues of the diagonalized photonic modes. The effective Hamiltonian that eliminates the
rapidly varying frequencies may be obtained following the recipe described in Sec. \ref{cyclic}. We quote the result below:
\beq
{\cal H}^{\hbox{eff}}  = H_{0} + \frac{\mathsf{g}^{2}}{2}\;\sum_{j = 0}^{N-1}
\frac{1}{\hat{\Delta}_{j}} \,\left(\mathsf{S}_{j j}^{z}\,\left(\mathsf{A}_{j}^{\dagger} \mathsf{A}_{j}
+ \mathsf{A}_{j} \mathsf{A}_{j}^{\dagger}\right)
+ \left(\mathsf{S}_{j}^{+} \mathsf{S}_{j}^{-} + \mathsf{S}_{j}^{-} \mathsf{S}_{j}^{+}\right) \mathbb{I}\right),
\label{HKR_K_large}
\eeq
where the detuning parameters for the photon eigenmodes read
$\hat{\Delta}_{j} = \varepsilon - \widehat{\Omega}_{j} = \delta - \kappa \,(N - 1 - 2 j)$. Towards obtaining the transmission of
single excitation quantum states we study the evolution of the coefficients $\mathsf{a}_{j}(t), \mathsf{b}_{j}(t)$ of the
localized excitations. In the present limit the equations (\ref{Kab_exp}) in conjunction with (\ref{alpha_beta}) produce
the time-dependent coefficients as
\beq
\mathsf{a}_{j}(t) = \sum_{k = 0}^{N-1}\; \mathcal{K}_{\mathsf{a}}(j, k; t)\;\mathsf{a}_{k}(0),\qquad
\mathsf{b}_{j}(t) = \sum_{k = 0}^{N-1}\; \mathcal{K}_{\mathsf{b}}(j, k; t)\;\mathsf{b}_{k}(0),
\label{Kab_evolve}
\eeq
where the kernels encoding the correlation functions read
\bea
\mathcal{K}_{\mathsf{a}}(j, k; t) &=&  \sum_{\ell = 0}^{N-1}\; \mathsf{U}_{j \ell}\, \mathsf{U}_{k \ell}\;
\exp\left(- i \left(\widehat{\Omega}_{\ell} - \frac{\mathsf{g}^{2}}{\hat{\Delta}_{\ell}}\right) t\right),\nn\\
\mathcal{K}_{\mathsf{b}}(j, k; t) &=& \sum_{\ell = 0}^{N-1}\; \mathsf{U}_{j \ell} \,\mathsf{U}_{k \ell}\;
\exp\left(- i \left(\varepsilon + \frac{\mathsf{g}^{2}}{\hat{\Delta}_{\ell}}\right)\,t\right).
\label{K_green}
\eea
Of particular significance is the case where the dispersive coupling between the atoms and the corresponding localized
cavity excitations is neglected: $\mathsf{g} = 0$. The evolution kernel of the localized photonic state reads as
\beq
\mathcal{K}_{\mathsf{a}}(j, k; t) = \exp \sl{- i (\Omega + \kappa (N - 1)) t}\; f_{j, k} (t)\qquad
f_{j, k}(t) = \sum_{\ell = 0}^{N-1}\; \mathsf{U}_{j \ell}\, \mathsf{U}_{k \ell}\;z^{\ell} \qquad
z = \exp(i 2 \kappa t).
\label{g0a}
\eeq
By employing a classical summation \cite{Bateman1953} formula of the hypergeometric series the time-dependent correlation between the
$j$-th and $k$-th sites may be expressed in a closed form :
\beq
f_{j, k}(t)= \frac{1}{2^{N - 1}}\;\sqrt{\binom{N - 1}{j}\binom{N - 1}{k}} (1-z)^{j+k}(1+z)^{N - 1-j-k}
{\ }_2F_1\left(\atop{-j,-k}{-(N - 1)} ; \frac{- 4z}{(1-z)^2}\right).
\label{Kfjk}
\eeq
In the context of a linear spin chain with the interaction determined by a Jacobi matrix such propagator of a single
excitation state was earlier obtained in \cite{CV2010}. Assuming that the initial state in (\ref{Kfjk}) is given by a photon in $0$-th
cavity: $\mathsf{a}_{j}(0) = \delta_{j 0}$, the
transmission of the photonic excitation to an arbitrary cavity is described by the evolution of the said coefficient as follows:
\beq
\mathsf{a}_{j}(t)= (- i)^{j}\, \sqrt{\binom{N - 1}{j}}\;\exp (- i \Omega t)\; \big(\sin (\kappa t)\big)^{j}\;
\big(\cos (\kappa t)\big)^{N - 1 -j}.
\label{KFN}
\eeq
It is evident from the above structure that in the dispersion-free $\mathsf{g} = 0$ case the amplitude of the photonic excitation
at the $(N - 1)$-th cavity is of unit magnitude $\left|\mathsf{a}_{j}(t)\right| = \delta_{j\; N -1}$
at the following predetermined times: $\{t = t_{n} \equiv \frac{(2 n + 1) \pi}{2 \kappa}| n = 0, 1, \ldots\}$.

\par

In the presence of the coupling  between the atom and the cavity photon $\mathsf{g} \neq 0$ it is not possible in general to obtain a closed
form expression for the time-evolution of the coefficient of the single photonic excitation state. To demonstrate that an exact transmission
for this case is not possible  in the present limit $\kappa\gg \delta, \mathsf{g}$ we proceed as follows. For the choice of initial state
$\mathsf{a}_{j}(0) = \delta_{j 0}$ the evolution kernels
given in (\ref{Kab_evolve}, \ref{K_green}) may be employed to express the magnitude of the excitation in the $(N - 1)$-th site as given below:
\beq
\left|\mathsf{a}_{N - 1}(t)\right| = \left|\sum_{\ell = 0}^{N-1}\; \mathsf{U}_{N-1 \;\ell}\, \mathsf{U}_{0\; \ell}\;
\exp \sq{i \Big(2 \kappa \ell + \frac{\mathsf{g}^{2}}{\hat{\Delta}_{\ell}}\Big) t}\right|.
\label{K_mag}
\eeq
The parity relation (\ref{KU_rel}) allows us to recast (\ref{K_mag}) as
\beq
\left|\mathsf{a}_{N - 1}(t)\right| = \left|\sum_{\ell = 0}^{N-1}\; \Big(\mathsf{U}_{0 \;\ell}\Big)^{2}\;
\exp \sl{i \sq{(2 n + 1) \pi \ell + \Big(2 \kappa \ell + \frac{\mathsf{g}^{2}}{\hat{\Delta}_{\ell}}\Big) t}}\right|\qquad n =0, 1, \ldots.
\eeq
The relative phase factors of the terms in the above summand do not, in general, identically reduce to zero for any choice of time.
Consequently, the magnitude of the excitation $\mathsf{a}_{N - 1}(t)$  obeys the inequality
\beq
\left|\mathsf{a}_{N - 1}(t)\right| \leq \left|\sum_{\ell = 0}^{N-1}\; \Big(\mathsf{U}_{0 \;\ell}\Big)^{2}\right| = 1.
\label{inequality}
\eeq
Only in the vanishing limit of the coupling $\mathsf{g} \rightarrow 0$ between the atom and the cavity photon, and also for large detuning limit
$\delta \gg \kappa, \mathsf{g}$ to be discussed next the excitation $\left|\mathsf{a}_{N - 1}(t)\right|$ equals its unit limiting value
for specific predetermined values of time.

\par

Proceeding parallel to the derivation in Sec. \ref{cyclic} the effective Hamiltonian in the limit of large detuning
parameter $\delta \gg \kappa, \mathsf{g}$  may be obtained. Up to the order $O(\mathsf{g}^{2})$ the effective Hamiltonian reads
\beq
{\cal H}^{\hbox{eff}}  = H_{0} + \frac{\mathsf{g}^{2}}{2}\,
\sB{\sum_{j, k = 0}^{N - 1}\,\frac{1}{\hat{\Delta}_{j}} \,\left(\mathsf{S}_{j k}^{z}\,\big(\mathsf{A}_{j} \mathsf{A}_{k}^{\dagger}
+ \mathsf{A}_{j}^{\dagger} \mathsf{A}_{k}\big)\right)
+ \sum_{j = 0}^{N - 1}\,\frac{1}{\hat{\Delta}_{j}} \,\left(\mathsf{S}_{j}^{+} \mathsf{S}_{j}^{-}
+ \mathsf{S}_{j}^{-} \mathsf{S}_{j}^{+}\right) \mathbb{I}}.
\label{KH_D_large}
\eeq
The propagation of single excitation states follows the description given in (\ref{Kab_evolve}, \ref{K_green}). To evaluate the propagator we
expand the exponent in  (\ref{K_green}) retaining terms up to the order
$O(\frac{\kappa}{\delta})$, and neglecting higher order terms in the said coefficient.  The kernel
$\mathcal{K}_{\mathsf{a}}(j, k; t)$ given in (\ref{K_green}) may now be evaluated exactly as before reproducing the result (\ref{g0a}, \ref{Kfjk})
with a {\it redefinition} of the constants:
\beq
\Omega \longrightarrow \Omega^{\prime} = \Omega\left(1 - \frac{\mathsf{g}^{2}}{\delta}\right),
\qquad \kappa \longrightarrow \kappa^{\prime} = \kappa \left(1 - \Big(\frac{\mathsf{g}}{\delta}\Big)^{2}\right).
\label{redef}
\eeq
We reproduce the kernel $\mathcal{K}_{\mathsf{a}}(j, k; t)$ as follows:
\beq
\mathcal{K}_{\mathsf{a}}(j, k; t) = \exp \sl{- i (\Omega^{\prime} + \kappa^{\prime} (N - 1)) t}\;
\sum_{\ell = 0}^{N-1}\; \mathsf{U}_{j \ell}\, \mathsf{U}_{k \ell}\;\big(z^{\prime}\big)^{\ell}, \qquad
z^{\prime} = \exp(i 2 \kappa^{\prime} t).
\label{Ka}
\eeq
Therefore, as in the dispersion free limit discussed in (\ref{KFN}),
the exact transmission of a single photon state from the $0$-th to $(N - 1)$-th cavity also takes place when in the
exponent of  the relative phases in the correlation function (\ref{K_green}) we retain
only the linear terms in the parameter $\frac{\kappa}{\delta}$. The time interval of exact propagation of the state now assumes the value
$t^{\prime}_{n} = \frac{(2 n + 1) \pi}{2 \kappa^{\prime}} \;\forall n = 0, 1, \ldots$. In the context of the approximation used here the
effect of a nonvanishing value of the coupling constant $\mathsf{g} \neq 0$ is to increase the time of transmission of the single
excitation state \,{\it i.\,e.} to decrease the velocity of propagation of the quantum state. Retaining terms of quadratic or higher order $O\big((\frac{\kappa}{\delta})^{2}\big)$ in evaluating the kernel
$\mathcal{K}_{\mathsf{a}}(j, k; t)$ defined in (\ref{K_green}) we observe that the exponent depends on the summation index at least
quadratically. Consequently, the summation no longer remains  of hypergeometric type, and therefore a general closed form expression is
not obtained by the present technique.

\par

Lastly we study the resonance limit $\varepsilon \approx \hat{\Omega}_{\ell} \Rightarrow \hat{\Delta}_{\ell} \approx 0$
for the $\ell$-th delocalized photon eigenstate in the presence of the parabolic coupling between the photons of adjacent cavities. We also
simultaneously assume that the following hierarchy of values holds: $\left|\frac{\hat{\Delta}_{\ell}}{\mathsf{g}}\right| \ll 1 ,\;
\left|\frac{\mathsf{g}}{\hat{\Delta}_{j}}\right| \ll 1 \;\forall j \neq \ell$. In contrast
to our description of the resonance scenario given in Sec. \ref{cyclic} the delocalized photon eigenstates here are not degenerate. In the resonance
limit the effective Hamiltonian has contributions of order $O(\mathsf{g})$ that couples atomic excitations with the photonic excitations:
\bea
{\cal H}^{\hbox{eff}}  &=& H_{0} + \mathsf{g} \; \left(\mathsf{S}^{+}_{\ell}\,\mathsf{A}_{\ell}
+ \mathsf{S}^{-}_{\ell}\,\mathsf{A}_{\ell}^{\dagger}\right)\nn\\
& & + \frac{\mathsf{g}^{2}}{2}\,
\sum_{\atop {j = 0} {j \neq \ell}}^{N - 1}\,\frac{1}{\hat{\Delta}_{j}} \,\sl{\mathsf{S}_{j j}^{z}\,\big(\mathsf{A}_{j} \mathsf{A}_{j}^{\dagger}
+ \mathsf{A}_{j}^{\dagger} \mathsf{A}_{j}\big)
+ \big(\mathsf{S}_{j}^{+} \mathsf{S}_{j}^{-} + \mathsf{S}_{j}^{-} \mathsf{S}_{j}^{+}\big) \mathbb{I}}.
\label{KH_res}
\eea
Assuming that the initial excitation is only atomic in nature: $\mathsf{a}_{j}(0) = 0 \;\forall j \in (0, 1,\ldots,  N - 1)$ the time-evolution of the
coefficients may be obtained via (\ref{Kab_exp}, \ref{alpha_beta}) as given below
\bea
\mathsf{a}_{j}(t) &=& - i\,\exp\sl{-i\left(\varepsilon - \frac{\Delta_{\ell}}{2}\right) t}\;
\sin\sl{\left(\mathsf{g} +\frac{\Delta_{\ell}^{2}}{8 \mathsf{g}}\right) t} \;
\sum_{k = 0}^{N - 1} \; \mathsf{U}_{j \ell} \,\mathsf{U}_{k \ell}\,\mathsf{b}_{k}(0),\nn\\
\mathsf{b}_{j}(t) &=& \sum_{k = 0}^{N - 1}\,\s{\sum_{\atop{n = 0}{n \neq \ell}}^{N - 1}\,
\,\mathsf{U}_{j n}\, \mathsf{U}_{k n}\,\exp\left(-i\left(\varepsilon + \frac{\mathsf{g}^{2}}{\Delta_{n}}\right) t\right)\nn\\
& & +  \mathsf{U}_{j \ell}\, \mathsf{U}_{k \ell}\;\exp\left(-i\left(\varepsilon - \frac{\Delta_{\ell}}{2}\right) t\right)\;
\cos\left(\left(\mathsf{g} +\frac{\Delta_{\ell}^{2}}{8 \mathsf{g}}\right) t\right)}\;\mathsf{b}_{k}(0).
\label{Koff_res-1}
\eea
For the exact resonance case $\Delta_{\ell} = 0$, and in the absence of dispersive effects
$\frac{\mathsf{g}}{\hat{\Delta}_{j}} \sim 0 \;\forall j \neq \ell$ the above evolution equations with the initial condition
$\mathsf{b}_{j}(0) = \delta_{j 0}$ read
\beq
\mathsf{a}_{j}(t) = - i \,\mathsf{U}_{0 \ell}\, \mathsf{U}_{j \ell}\exp(-i \varepsilon t)\,\sin(\mathsf{g} t),\qquad
\mathsf{b}_{j}(t) = \exp(-i \varepsilon t)\,\sl{\delta_{j 0}
- 2\,\mathsf{U}_{0 \ell} \,\mathsf{U}_{j \ell}\sin^{2}\left(\frac{\mathsf{g} t}{2}\right)}.
\label{Kab_res}
\eeq
In particular the excitation of the $(N - 1)$-th atom reads
\beq
\mathsf{b}_{N-1}(t) = - 2\,\exp(-i \varepsilon t)\,\mathsf{U}_{0 \ell} \,\mathsf{U}_{N-1\, \ell}\,\sin^{2}\left(\frac{\mathsf{g} t}{2}\right)
= 2\,(-1)^{\ell + 1}\,\exp(-i \varepsilon t)\,\left(\mathsf{U}_{0 \ell}\right)^{2} \,\sin^{2}\left(\frac{\mathsf{g} t}{2}\right),
\label{b_3}
\eeq
where in the second equality we have used the parity relation (\ref{KU_rel}).
As it may be observed from the construction (\ref{KU_def}, \ref{K-tilde}) of the orthogonal matrix $\mathsf{U}$ that in the exact resonance
regime for the parabolic coupling a perfect transmission of the quantum state is realized for the length of the array $N = 3$, and for the
resonance mode $\ell = 1$ at the time $\frac{(2 n + 1) \pi}{\mathsf{g}} \; \forall n \in (0, 1, \ldots)$.
\section{Conclusion}
\label{conclusion}
Here we have considered one dimensional array of optical cavities with a nearest-neighbor hopping interaction of the
cavity photons described by Jaynes-Cummings-Hubbard model. In particular we have studied the time evolution of single
excitation states in such chains of coupled cavities as models for quantum communication. Two different choices of
coupling coefficients linking photons of adjacent cavities have been investigated. Employing delocalized collective
photonic and atomic modes we first study a translation invariant closed chain of an arbitrary number of $N$ identical
cavities with uniform coupling constant. Following this we have considered a linear chain of cavities where the
non-uniform parabolic hopping term of the Hamiltonian is related to the tridiagonal Jacobi matrix associated with the Krawtchouk
polynomials. For both of these cases we obtain the effective Hamiltonian in various simplifying limits. For dominant value of
photonic hopping parameter, as well as for large detuning parameter between the atomic and the photonic frequencies the
excitations are transferred between, say, the atoms without populating the field modes. Expressed via the delocalized atomic modes the
spin-spin interaction term in the effective Hamiltonian is diagonalized. The atoms experience a Stark shift dependent upon
the population of the field modes. In the case of resonance between the atoms and a delocalized photonic mode propagation of
excitation between the atoms requires intermediate excitation of a photonic mode. In the case of non-uniform parabolic coupling
between the photons of adjacent cavities the transmission of single-excitation states is exact for the limiting value
$\mathsf{g} \rightarrow 0$ of the atom-cavity photon coupling constant. However in the large detuning limit where we retain
only the linear terms in the parameter $\frac{\kappa}{\delta}$ the propagation of the single excitation states remains exact
with an increase in the time of transmission. This is observed in our evaluation of the time-dependent
correlation function of the one-excitation states.

\par

The analysis developed here may have applications in certain contexts. The idea of having optical lattice systems with pre-engineered
coupling constants between individual lattice sites may give rise to interesting physical situations. In particular the resonance situation
where mixing of atomic and field modes is realized merits attention. In the present model it may be possible to generate multipartite entangled states of
polaritonic qubits and study the time variation in the entanglement in the presence of dissipative atom-cavity photon coupling. Moreover,
in the context of one dimensional spin chains it has recently been observed \cite{JV2010} that coupling between lattice sites determined by
the Jacobi matrices of $q$-deformed Krawtchouk polynomials ensure perfect transmission of quantum states. It should be useful to study the
problem in the context of Jaynes-Cummings-Hubbard lattices where coupled atomic and photonic modes are considered. The delocalized collective
modes discussed here may allow easy extraction of the effective Hamiltonian at the desired perturbative level. Another possible application
of the present method may lie in the quantum phase transitions in coupled array of atom-cavity photon systems \cite{MFIS2009}. Specific
pre-engineered coupling between the cavity sites may give rise to desirable properties for a quantum transition between the Mott insulator
and the superfluid states.

\end{document}